\documentclass[aps,pra,superscriptaddress,amsmath,amssymb,showpacs]{revtex4-2}
\usepackage[english]{babel}
\usepackage[cp866]{inputenc}
\usepackage{amssymb,bm}
\usepackage{graphicx}
\usepackage{amsmath}
\usepackage{color}
\usepackage[colorlinks=true,citecolor=blue,urlcolor=blue]{hyperref}
\begin{document}

	\title{Effects of isospin violation in the $e^+e^- \rightarrow B^{(*)}\bar B^{(*)}$ cross sections}
\author{A.E. Bondar}
\email{A.E. Bondar@inp.nsk.su}
\affiliation{Budker Institute of Nuclear Physics of SB RAS, 630090 Novosibirsk, Russia}
\affiliation{Novosibirsk State University, 630090 Novosibirsk, Russia}
	\author{A.I. Milstein}
	\email{A.I.Milstein@inp.nsk.su}
	\affiliation{Budker Institute of Nuclear Physics of SB RAS, 630090 Novosibirsk, Russia}
	\affiliation{Novosibirsk State University, 630090 Novosibirsk, Russia}
		\author{R.V. Mizuk}
	\email{roman.miziuk@gmail.com}
	\affiliation{P.N. Lebedev Physical Institute, RU-117924 Moscow, Russia}
	\author{S.G. Salnikov}
	\email{S.G.Salnikov@inp.nsk.su}
	\affiliation{Budker Institute of Nuclear Physics of SB RAS, 630090 Novosibirsk, Russia}
	\affiliation{Novosibirsk State University, 630090 Novosibirsk, Russia}
		\date{\today}

	\begin{abstract}
		Model estimates are obtained for the
		influence of Coulomb effects on the ratio of the cross sections for the production of charged and neutral $B\bar B$ and $B^*\bar B^*$ pairs in $e^+e^-$ annihilation. It is shown that the difference between the masses of charged and neutral mesons obtained under the assumption that this ratio is constant on a scale of the order of the beam energy spread can differ from the true one by $\delta M \sim 0.03\,\mbox{MeV}$ at the energy of $\Upsilon (5S)$ and by $\delta M \sim 0.4\,\mbox{MeV}$ at the energy of $\Upsilon (4S)$. Thus, the errors given in the PDG for the difference between the masses of charged and neutral $B$ mesons, based on the results obtained at the energy of $\Upsilon (4S)$, are strongly underestimated. Similar measurements at the energy of $\Upsilon (5S)$ will have an order of magnitude smaller systematic shift for the mass difference. This circumstance should be taken into account when planning future experiments at the $B$ factory in KEK.
	\end{abstract}
	\maketitle

\section{Introduction}
At present, there is a large amount of experimental data on the production of resonances in the baryon-antibaryon and meson-antimeson systems in the energy region near the reaction threshold $M_{th}$. For these resonances, the difference between the resonance mass $M_R$ and $M_{th}$ is comparable to the resonance width~$\Gamma_R$. As a result, the shape of the spectral line, i.e., the energy dependence of the resonance production cross section, has a very nontrivial form. Experimental and theoretical studies of near-threshold resonances provide information about the properties of the strong interaction in the energy region where perturbation theory is not applicable. In addition, the relative velocity of interacting hadrons for near-threshold resonances is small. Therefore, the shape and position of the resonance can be affected by the Coulomb interaction of charged particles and the mass difference between charged and neutral particles, that is, the effects of violation of isotopic invariance. In a recent paper \cite{MS2021}, the Coulomb effects were discussed in the $\Upsilon (4S) \rightarrow B^0\bar B^0$ and $\Upsilon (4S) \rightarrow B^+B^-$ decays, in which the near-threshold resonances are observed, with $M_R-M_{th}= 20.1\pm 1.4\,\mbox{MeV}$ for neutral mesons and $ 20.7 \pm 1.4 \, \mbox{MeV}$ for charged mesons, $\Gamma_R= 20.5 \pm 2.5 \,\mbox{MeV}$. In particular, in \cite{MS2021} the ratio of the cross section for the production of charged mesons to the cross section for the production of neutral mesons, $\sigma_c/\sigma_n$, was studied. We used experimental data from \cite{CLEO2001,CLEO2002,BABAR2004,BABAR2009,Dong2020,Belle2021}. In early theoretical papers \cite{Marciano1990, Lepage1990,Eichten1990,Kaizer2003,Voloshin2003,Voloshin2005} completely different approaches to estimating the magnitude of the Coulomb effects in $\Upsilon (4S) \rightarrow B\bar B$ decay with completely different predictions were used. The measurement of the line shape in the $\Upsilon (4S) \rightarrow B\bar B$ decay~\cite{BABAR2009, Dong2020} made it possible to obtain a significant progress in understanding the role of the Coulomb effects. Unfortunately, the line shape was determined only for the sum $\sigma_c+\sigma_n$, and the experimental value for the ratio $\sigma_c/\sigma_n\sim 1.06\,$ given in the PDG~\cite{ParticleDataGroup:2020ssz} was obtained only at the maximum of the total cross section $\sigma_c+\sigma_n\,$.

Usually, when estimating the Coulomb effects, the factorization hypothesis is used, according to which
$$\sigma_c/\sigma_n=C=t/[1-\exp(-t)]\,,\quad t=\pi\alpha/v\,,$$
where $C$ is the Sommerfeld-Gamow-Sakharov factor, $v$ is the $B$-meson velocity, $\alpha$ is the fine structure constant, $\hbar = c = 1$. We have $t\approx 0.37$ and $\sigma_c/\sigma_n\sim 1.2\,$, which is noticeably larger than the result given in the PDG. This discrepancy has been explained in
\cite{MS2021}, in which it was shown that the factorization hypothesis does not work for near-threshold resonances, and the position of the peak for charged $B$ mesons noticeably differs from peak position for neutral $B$ mesons even if the masses of these mesons are equal. As a result, the position of the $\sigma_c+\sigma_n$ maximum turned out to be between the $\sigma_c$ and $\sigma_n$ maxima, which led to a small value of $(\sigma_c/\sigma_n-1)\,$. In addition, it was shown in \cite{MS2021} that the Coulomb effects lead to a rather nontrivial dependence of the ratio $\sigma_c/\sigma_n$ on energy.

Accounting for violation of isotopic invariance during the production of mesons with heavy quarks in $e^+e^-$ annihilation can be important for determining the mass difference between charged and neutral particles from experimental data. At present, the mass difference of charmed mesons is known with much higher accuracy than that of beautiful mesons. In the case of pseudoscalar $B$
mesons, the world average value of the mass difference between
charged and neutral mesons is equal to $M_\mathrm{B^0}-M_\mathrm{B^+} =
0.32\pm0.05\,{\mathrm{MeV}}$~\cite{ParticleDataGroup:2020ssz}.
However, this result is mainly determined by the BaBar measurement at $\Upsilon (4S)$ resonance, $M_\mathrm{B^0}-M_\mathrm{B^+} = 0.33\pm0.05\pm0.03\,{\mathrm{MeV}}$~\cite{BaBar:2008ikz}, and was obtained using the assumption that the ratio of the production cross sections
of charged and neutral pairs of $B$ mesons is independent  of energy
near the maximum of $\Upsilon (4S)$ (except for the phase volume correction ($p_{B^+}/p_{B^0})^3$).
Estimates show that such an assumption may not be true. The point is that the ratio of the
production cross sections for charged and neutral $B$
mesons can change by tens of percent when the energy of ${e^+e^-}$ pair changes by $10\,\mathrm{MeV}$ near the maximum of $\Upsilon (4S)$~\cite{MS2021}. The difference in the energy derivatives of the ${B^+}$ and ${B^0}$ production cross sections will lead to different average energies of ${B^+}$ and ${B^0}$ mesons, the difference of which from the average beam energy is proportional to the energy derivative of the cross section and the square of the energy spread of the colliding beams. Neglecting this effect will lead to an incorrect extraction of the $B$-meson masses from the experimental data. Assuming an energy spread of $5\,{\mathrm{MeV}}$, the systematic error in the splitting of the masses of charged and neutral $B$ mesons in the BaBar measurement can reach $0.2\div 0.4\,\mathrm{MeV}$. The final conclusion about the reliability of this result can only be made after the
experimental measurement of the energy dependence of the ratio of
cross sections for the production of charged and neutral mesons near the
maximum of $\Upsilon (4S)$. To do this, it is necessary to measure the ratio of the yields of charged  and neutral $B$ mesons in a certain neighborhood near the maximum of the cross section. Since the stability of the beam energy in the Belle and BaBar experiments was not ideal, in principle it is possible to obtain from the experimental data, with some accuracy, the value of the energy derivative of the ratio of the yields of charged and neutral   $B$ mesons and to make a correction to the value of the mass splitting of the  pseudoscalar beauty mesons.

As regards the isotopic splitting of $B^*$ mesons, currently there is only an upper bound, $|M_\mathrm{B^{*0}} - M_\mathrm{B^{*+}}| < 6\,\mathrm{MeV}$, obtained in 1995 in the DELPHI~\cite{DELPHI:1995hfn} experiment. Since Belle has a large amount of experimental data at the $\Upsilon (5S)$ meson energy, the mass difference between the charged and neutral $B^*$ mesons can be measured with much better accuracy than in the DELPHI experiment. To estimate the systematic error in such a measurement, it is very useful to estimate the influence of isotopic invariance violation on the cross section for the production of $B^*$ mesons at the energy of $\Upsilon (5S)$.

In a recent work by Belle~\cite{Belle2021}, the energy dependence of the cross sections for the processes $e^+e^- \rightarrow B\bar B,\, B\bar B^*\,, B^*\bar B^* \, $ was measured. In all these processes pairs of mesons are produced in a state with an orbital momentum $l=1$, and near-threshold resonances are observed. In our paper, using the experimental data~\cite{Belle2021} and the theoretical approach~\cite{MS2021}, we study the influence of the Coulomb interaction on the energy dependence of the $e^+e^- \rightarrow B\bar B,\, B^*\bar B^*$ cross sections in the charged and neutral modes. The shape of the resonances is usually described within the Breit-Wigner approximation, which is not applicable to the near-threshold resonances. In the latter case, the Flatt\'{e} formula \cite{flatte76} is used. It is more convenient, however, to apply a different approach used in \cite{MS2021}. The point is that the interaction potentials of $ B $ mesons are poorly known. However, many isotopically invariant potentials can be found that describe the shape of the resonance found experimentally.A situation is similar to the potential models of heavy quarkonia, for which there is a large number of potentials that well describe the spectrum of quarkonia and their lepton widths. In addition, among these potentials there are those for which it is easy to carry out numerical and analytical calculations. By adding the Coulomb interaction to such potentials, one can study the influence of the Coulomb interaction on the shape of the resonances. Naturally, at the end it is necessary to verify that the Coulomb effects weakly depend on the magnitude of the model strong interaction potential.

\section{Theoretical approach}
The general approach to describing near-threshold resonances was formulated in Refs. \cite{DMS2014, DMS2016, MS2018}. In the special case of $B\bar B$ pair production in a state with an orbital momentum $l=1$, the method is described in Ref.~\cite{MS2021}, where it is shown that the charge exchange potential $ V_{ex}(r)$ has small effect on the shape of the spectral line. This potential determines the transition amplitudes $B^+B^-\leftrightarrow B^0\bar B^0$. If we neglect the potential $V_{ex}(r)$, then to calculate the cross sections considered in the present work, it is necessary to find the radial part $U^{(n)}(r)$ of the wave function of the pair $B^{0(*)}\bar B^{0(*)}$ and the radial part $U^{(c)}(r)$ of the wave function of the pair $B^{+(*)}\bar B^{-(*)}$. These functions satisfy the equations
\begin{align}\label{wf}
&\left[\dfrac{p_r^2}{M_B}+
\dfrac{2}{M_Br^2}+V(r)+2\Delta-E \right]U^{(n)}(r)=0\,,\nonumber\\
& \left[\dfrac{p_r^2}{M_B}+
\dfrac{2}{M_Br^2}+V(r)-\dfrac{\alpha}{r}-E \right]U^{(c)}(r)=0\,.
\end{align}
Here $V(r)$ is the model potential, $-p_r^2$ is the radial part of the Laplacian, $M_B=(M_{B^0}+M_{B^+})/2$, and $\Delta= M_{B^0}-M_{B^+}$, the energy $E$ is measured from the threshold for the production of a pair of charged mesons.
Solutions $U^{(c)}(r)$ and $U^{(n)}(r)$ have the following asymptotics at large distances
\begin{align}
& U^{(c)}(r)=\frac{1}{2ik_cr}\left[S_c\chi_{c}^{+}(k_c,r)-\chi_{c}^{-}(k_c,r)\right],\nonumber \\
& U^{(n)}(r)=\frac{1}{2ik_nr}\left[S_n\chi_{n}^{+}(k_n,r)-\chi_{n}^{-}(k_n,r)\right],
\end{align}
Here $S_{c,n}$ are some functions of energy, $k_c=\sqrt{M_BE}$, $k_n=\sqrt{M_B(E-2\Delta)}$,
\begin{align}
& \chi_{c}^{\pm}(k,r)=\exp\left\{\pm i\left[kr- \pi/2+\eta_k\ln(2kr)+\delta_k\right]\right\},\nonumber \\
& \chi_{n}^{\pm}(k,r)=\exp\left[\vphantom{\bigl(\bigr)}\pm i\left(kr- \pi/2\right)\right],\nonumber \\
& \delta_k=\frac{i}{2}\ln\frac{\Gamma\left(2+i\eta_k\right)}{\Gamma\left(2-i\eta_k\right)}\,,\qquad
\eta_k=\frac{M_B\alpha}{2k}\,,
\end{align}
$\Gamma(x)$ is the Euler $\Gamma$ function.

The cross sections $\sigma_c$ and $\sigma_n$ for near-threshold resonances production in the processes $e^+e^- \rightarrow B^{+(*)}\bar B^{-(*)}$ and $e^+e^- \rightarrow B^{0(*)}\bar B^{0(*)}$ are equal, respectively,
 \begin{align}\label{3W}
&\sigma_c=N\,k_c\,\left|\dfrac{\partial}{\partial r}\,U^{(c)}(0)\right|^2\,,\quad
\sigma_n=N\,k_n\,\left|\dfrac{\partial}{\partial r}U^{(n)}(0)\right|^2\,,
\end{align}
where $N$ is some constant. The simplest model potential, which nevertheless makes it possible to describe well the shape of the near-threshold resonances, is $V(r)=-V_0\,\theta(a-r)$, where $V_0$ and $a$ are some parameters that depend from the considered resonance, $\theta(x)$ is the Heaviside function. For this potential, it is easy to find analytical solutions that simplify the analysis of the effects associated with the violation of isotopic invariance. We have
\begin{align}\label{wcnfinal}
&\sigma_c=b\, \dfrac{k_c\,q^2}{M_B^3}\,\Bigg|\dfrac{q\,C_q}{k_c\,H^{+}{}'(k_c, k_ca)\,{\cal F}(qa) -q\, H^+(k_c,k_ca)\,{\cal F}'(qa)}\Bigg|^2\,,\nonumber\\
&\sigma_n=b\, \dfrac{k_n\,q^2}{M_B^3}\,\Bigg|\dfrac{q\,}{k_n\,h^{+}{}'(k_na)\,f(qa) -q\, h^+(k_na)\,f'(qa)}\Bigg|^2\,.
\end{align}
Here $Z'(x)\equiv \partial Z(x)/\partial x$ and the following notations are used
\begin{align}
&H^+(k,x)=4i\exp[ix +i\delta_k-\pi\eta_k/2]\,x^2\,{\cal U}(2-i\eta_k,4,\,-2ix)\,,\nonumber\\
&{\cal F}(x)=\dfrac{C_q}{3}x^2 \,e^{-ix}\,F(i\eta_q+2,4,\,2ix)\,,\nonumber\\
&C_q=\sqrt{\dfrac{2\pi\eta_q\,(1+\eta_q^2)}{1-\exp(-2\pi\eta_q)}}\,,\quad q=\sqrt{M_B(E+V_0)}\,,\nonumber\\
&h^+(x)=\left(\dfrac{1}{x}-i\right)\,e^{ix}\,,\quad
f(x)=\dfrac{\sin x}{x}-\cos x\,.
\end{align}
Here $F(b,c,z)$ is a confluent hypergeometric function of the first kind, ${\cal U}(b,c,z)$ is a confluent hypergeometric function of the second kind, ${\cal U}(b,c,z) \rightarrow 1$ for $|z|\to\infty$. The constant $b$ in \eqref{wcnfinal} is chosen so that the cross sections of the processes $e^+e^-\to B\bar B$ coincide with the experimental data. The observed near-threshold resonances correspond to low-lying virtual levels in the $p$ wave, i.e., $V_0$ is much larger than the energy of the virtual level. Therefore, the potential $ V_0 $ can be chosen in the form
\begin{equation}\label{V0}
V_0=\dfrac{(n\pi)^2}{M_Ba^2} -\tilde{E}_R\,,\quad n=3,4,5,6,\dots\,,
\end{equation}
where $\tilde{E}_R$ is a parameter close to the value of the
resonance energy $E_R$. It weakly depends on $a$ and is almost independent of $n$.
The expression \eqref{wcnfinal} is exact within the model and very convenient for analyzing various effects. We are convinced that qualitatively our predictions correspond to the real experimental situation.

\section{Discussion of the results}

\begin{figure}[t]

	\includegraphics[totalheight=5.2cm]{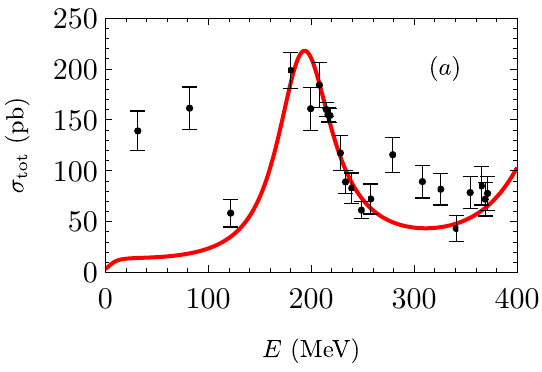}\hfill{}\includegraphics[totalheight=5.2cm]{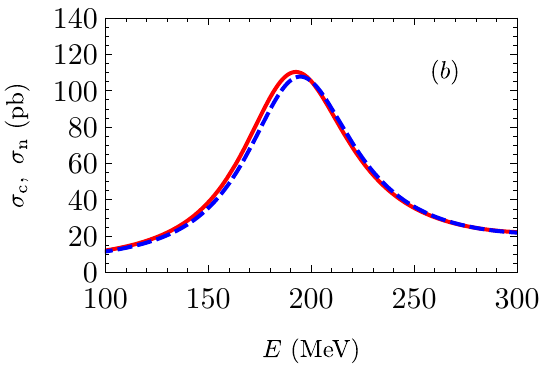}

	\includegraphics[totalheight=5.2cm]{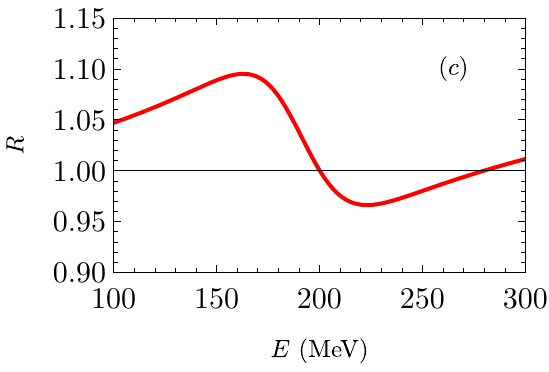}\hfill{}\includegraphics[totalheight=5.2cm]{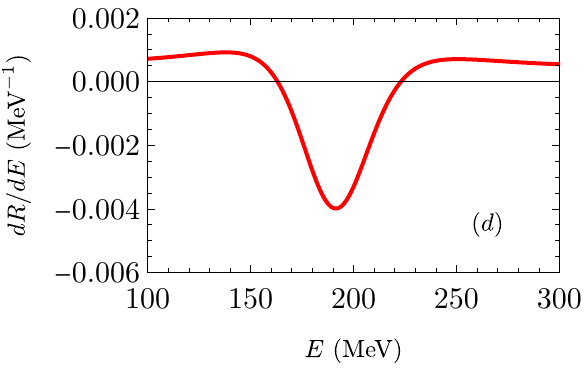}

	\caption{Cross sections of $B^{*}\bar{B}^{*}$ meson production for $a=2.5\,\text{fm}$,
		$V_0=1198\,\text{MeV}$.
		(a)~Energy dependence of $\sigma_\mathrm{tot}=\sigma_c+\sigma_n$.
		(b)~Energy dependence of $\sigma_c$ (solid curve) and $\sigma_n$ (dashed curve) near the $\Upsilon(5S)$ resonance.
		(c)~Energy dependence of the ratio $R=\sigma_{c}/\sigma_{n}$ near the $\Upsilon(5S)$ resonance.
		(d)~Energy dependence of $dR/dE$ near the $\Upsilon(5S)$ resonance.
		Experimental data are from Ref.~\cite{Belle2021}.}

	\label{BsBs}
\end{figure}

\begin{figure}[t]
	\includegraphics[totalheight=5.2cm]{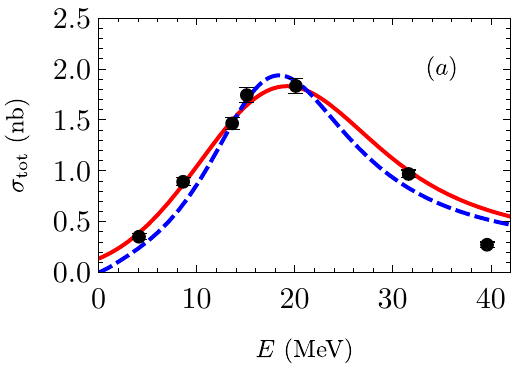}\hfill{}\includegraphics[totalheight=5.2cm]{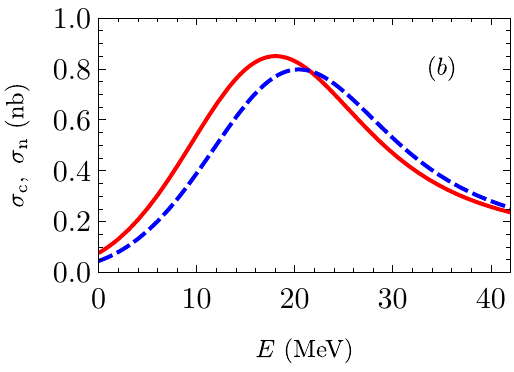}

	\includegraphics[totalheight=5.2cm]{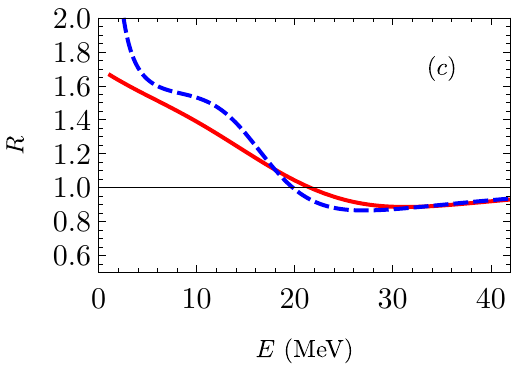}\hfill{}\includegraphics[totalheight=5.2cm]{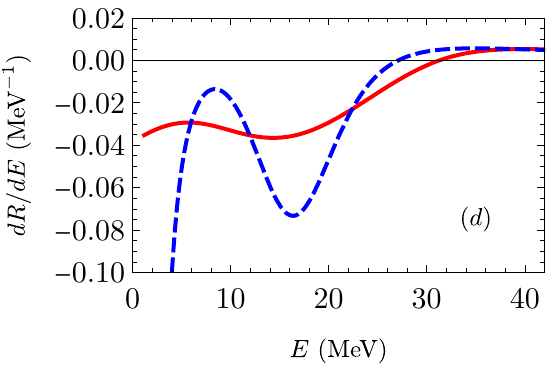}

	\caption{Cross sections of $B\bar{B}$ meson production for $a=2.5\,\text{fm}$,
		$V_0=269\,\text{MeV}$.
		(a)~Energy dependence of $\sigma_\mathrm{tot}=\sigma_c+\sigma_n$ with (solid curve) and without (dashed curve) averaging over the beam energy spread.
		(b)~Energy dependence of $\sigma_c$ (solid curve) and $\sigma_n$ (dashed curve)
		with averaging over the beam energy spread.
		(c)~Energy dependence of $R=\sigma_{c}/\sigma_{n}$ with (solid curve) and without (dashed curve) averaging over the beam energy spread.
		(d)~Energy dependence of $dR/dE$ with (solid curve) and without (dashed curve) averaging over the beam energy spread.
		Experimental data with the radiative corrections taken into account are from Ref.~\cite{Dong2020}.}
	\label{BB}
\end{figure}

Let us consider the processes $e^+e^- \rightarrow B^{*}\bar B^{*}$ near the $\Upsilon (5S)$ resonance, which corresponds to the energy $E \approx 200$~MeV. Experimental data for the sum $\sigma_c +\sigma_n $ were obtained in \cite{Belle2021}. The parameters of our model were chosen to fit the experimental data in the energy region $100\div 250\,\mathrm{MeV}$.
These parameters turned out to be $a=2.5\,{\text{fm}}$,
$V_0=1198\,{\text{MeV}}$. In principle, when comparing model predictions with the experimental data, it is necessary to take into account the energy spread $\Delta_0$ in the $e^+e^-$ beams. To do this, one should average the cross sections according to the formula
\begin{equation}
\langle \sigma(E)\rangle=\int_0^\infty \sigma(E')\exp\left[-\dfrac{(E-E')^2}{2\Delta_0^2}\right] \dfrac{dE'}{\sqrt{2\pi}\,\Delta_0}\,.
\nonumber
\end{equation}
In the Belle and BaBar experiments $\Delta_0\approx5\,\mathrm{MeV}$. It turned out that such averaging is essential in the case of $B$-meson production and not essential in the case of $B^*$ mesons.
Figure~\ref{BsBs} shows the experimental data and model predictions for $\sigma_c +\sigma_n $, the $\sigma_c$ and $\sigma_n $ cross sections, their ratio, and the energy derivative of the ratio of the cross sections. It can be seen that the model with the chosen parameters describes well the experimental data in the energy region $100<E<250$~MeV. It is in this region where Belle has the most of the data for the $e^+e^- \rightarrow B^{*}\bar B^{*}$ process. Therefore, these data can be used to determine the mass difference between charged and neutral $B^*$ mesons.
To do this, it is necessary to estimate the theoretical uncertainty in this difference associated with the Coulomb interaction of mesons. As can be seen from Fig.~\ref{BsBs}(b), the energy dependence of the cross sections $\sigma_c$ and $\sigma_n$ are close to each other. The difference of these cross sections in the case of $e^+e^- \rightarrow B^{*}\bar B^{*}$ is noticeably smaller than in the case of $e^+e^- \rightarrow B \bar B$ at $\Upsilon (4S)$, see Fig.~\ref{BB}. This is due to the fact that the relative velocity of $B$ mesons at $\Upsilon (4S)$ is much less than in the case of the production of $B^*$ mesons at $\Upsilon (5S)$. As a result, the difference from unity of the ratio $\sigma_c/\sigma_n$ (Fig.~\ref{BsBs}(c)) and the energy derivative of this ratio for $B^* \bar{B^*}$ (Fig.~\ref{BsBs}(d)) is much less than for $B\bar B$ (Figs.~\ref{BB}(c,d)).

The results obtained allow us to estimate the systematic shift $\delta M$ in the difference between the masses of charged and neutral $B^{(*)}$ mesons that arises in the experimental analysis if the energy dependence of the cross section ratio ${\sigma_c}/{\sigma_n}$ is neglected. The extraction of the difference between the masses of charged and neutral mesons in the experiments with colliding $e^+e^-$ beams was based on measuring the difference in the momenta of the reconstructed $B$ mesons under the assumption that the total energies of mesons are equal to the average energy of the electrons and positrons in the beams. This assumption is valid only in the case of a constant production cross section in the energy spread interval $\Delta_0$. In fact, we see that the Coulomb interaction violates this assumption for the cross section ratio ${\sigma_c}/{\sigma_n}$. Assuming that this ratio depends linearly on the energy $E$ in an interval of order $\Delta_0$, we arrive at the formula
\begin{equation}
\delta M\approx\frac{\Delta_0^2}{2}\,\frac{\partial}{\partial E}\,\frac{\sigma_c}{\sigma_n},
\label{DM}
\end{equation}
where the ratio $\sigma_c/\sigma_n$ should be normalized to unity at the nominal $e^+e^-$ energy. Using this formula, we obtain the expected systematic shift $\delta M\approx 0.03\,\mathrm{MeV}$ in the case of the production of $B^*$ mesons at $\Upsilon (5S)$ and $\delta M\approx 0.4\,\mathrm{MeV}$ for $B$ mesons at $\Upsilon (4S)$. As already noted, several different values of the potential parameters can be chosen, which describe the experimental data quite well. We have verified that the estimate of the systematic shift of the mass difference is stable with respect to the choice of specific potential values. As the energies of pseudoscalar $B$~mesons at $\Upsilon(5S)$ resonance are close to that of $B^*$ mesons, the energy derivative of the ratio $\sigma_c/\sigma_n$ for $B$~mesons at $\Upsilon(5S)$ is also of the order $10^{-3}\,\mathrm{MeV}^{-1}$. Therefore, we expect the systematic shift to be $\delta M\sim 0.03\,\mathrm{MeV}$ in the case of $B$-meson production at $\Upsilon (5S)$ too.

\section{Conclusion}
Within our model, we have obtained the estimates for
influence of Coulomb effects on the ratio $\sigma_{c}/\sigma_{n}$ for the production of $B\bar B$ and $B^*\bar B^*$ pairs. Using these estimates, we have shown that the difference between the masses of charged and neutral mesons obtained under the assumption that the ratio $\sigma_{c}/\sigma_{n}$ is constant over the interval $\Delta_0$ can differ from reality by $\delta M \sim 0.03\,\mathrm{MeV}$ at the energy of $\Upsilon (5S)$ and by $\delta M \sim 0.4\,\mathrm{MeV}$ at the energy $\Upsilon (4S)$. Thus, the errors given in the PDG for the difference between the masses of charged and neutral $B$ mesons based on the results obtained in Ref.~\cite{BaBar:2008ikz} at the energy of $\Upsilon (4S)$ are greatly underestimated. It follows from our results that similar measurements at the energy of $\Upsilon (5S)$ will have an order of magnitude smaller systematic shift for the mass difference. This circumstance should be taken into account when planning future experiments at the $B$ factory in KEK.

%\section*{Acknowledgments}
% We are grateful to A.E.~Bondar and R.V.~Mizuk for useful discussions.

 \end{document}